\documentclass[10pt,journal,compsoc]{IEEEtran}

\usepackage{epsfig}
\usepackage[cmex10]{amsmath}
\usepackage{amsfonts}
\usepackage{mathrsfs}
\usepackage{alltt}
\usepackage{url}

\usepackage{atbeginend}
\BeforeBegin{alltt}{\small}
\AfterEnd{alltt}{\normalsize}

\usepackage[english]{babel}

\begin{document}

\title{Redefining The Query Optimization Process}

\author{Kristian~F.~D.~Rietveld
        and~Harry~A.~G.~Wijshoff%
\IEEEcompsocitemizethanks{%
\IEEEcompsocthanksitem Originally submitted to IEEE Transactions on
Knowledge and Data Engineering, January 2014.%
\IEEEcompsocthanksitem K.~F.~D.~Rietveld and H.~A.~G.~Wijshoff
are with the Leiden Institute for Advanced Computer Science, Leiden
University, Leiden, The Netherlands.}%
\thanks{}}

\IEEEcompsoctitleabstractindextext{
\begin{abstract}
Traditionally, query optimizers have been designed for computer systems that
share a common architecture, consisting of a CPU, main memory and disk
subsystem. The efficiency of query optimizers and their successful
employment relied on the fact that these architectures basically stayed the
same over the last decades. However, recently the performance increase of
serial instruction execution has stagnated. As a consequence, computer
architectures have started to diversify. Combined with the fact that the
size of main memories has significantly increased it becomes more
important to exploit intrinsic internal features of computer systems (data
coherence mechanisms, TLB and instruction cache performance, among others)
rather than mainly focusing on minimizing disk I/O. Time has come for a
re-evaluation of how (traditional) query optimization is implemented.
Query optimizers must be able to cope with disruptive
advances in computer architecture and employ methods to optimize for
architecture-specific features. In this paper, the query optimization
process is redefined so that compiler optimizations are taken into the game
much earlier than traditionally is done and employ
compiler optimizations as the driving force in query optimization.
This new generation of query optimizers will be capable of optimizing
queries to perform significantly better than contemporary
state-of-the-art query optimizers.
\end{abstract}

%\category{D.3.4}{Programming Languages}{Processors}[Compilers, Optimization]
%\category{H.2.4}{Database Management}{Systems}[Query Processing, Relational Databases]

\begin{IEEEkeywords}
Optimizing Compilers, Relational Databases, Query Processing, Query
Optimization, Program Transformation, Intermediate Representation
\end{IEEEkeywords}}

\maketitle

%% Section: Introduction
\section{Introduction}
In the last decade, the increase of single-core performance of CPUs has
stagnated. Up till now, database systems could increase their performance to
a large extent by relying on these increases in single-core performance. This
is no longer the case. In the foreseeable future, the overall improvement of
(multi-core) CPU performance will stagnate as well, as it is no longer
practical to put more cores in a single CPU due to cache coherence issues
and the physical limitations for transistor sizes. Since the needs for data
processing grow exponentially, this would also result in exponential growth
(of the size) of hardware platforms. This is not attainable for the near
future. Therefore, the performance of database systems will become more
critical than it is today and require a re-evaluation of how database
systems should be designed.

An additional consequence of the stagnation of single-core performance is
that computer architectures will start to diversify even more than they are
currently doing and this will result in more disruptive changes of computer
architectures. Disk subsystems might diversify by introducing flash storage
in a traditional disk subsystem, omitting mechanical disks or providing data
to be processed through a high-speed network connection. In the future,
different CPU architectures will emerge with varying numbers of cores with
different capabilities, differing cache architectures and coherence schemes
combining software mechanisms with hardware-based schemes. Traditionally,
query optimizers have been designed for generations of computer systems
assuming that changes in architecture are non-disruptive.  Due to the
growing diversity of computer architectures in the future this assumption no
longer holds and query optimizers will need significant adaptations to deal
with future advances in computer architecture.

Compiler technology forms a crucial component of future releases of
computer architectures. New computer architectures will be disruptive,
requiring compiler heuristics to be extensively re-tuned for
achieving acceptable performance. New features of such architectures will be
controlled by software mechanisms, configuring these mechanisms
according to the characteristics of the software that will be run.
Furthermore, for specific architectural features, specific compiler
transformations might be developed. The state of art in compiler technology
for future systems will therefore be of crucial importance for successful
deployment of these systems.

For database systems these developments will mean that performance
can no longer be taken for granted and that database systems will become
more and more dependent on specific architectural features as well as
compiler support for these features. In general one could assume that
specific dedicated hardware would be developed for database systems in which
specific database management backends will be implemented on top of these
architectural features.  The other solution would be to integrate database
systems more and more with compiler technology. Although the first solution
will certainly form a feasible solution, in this paper we would like to
emphasize that the integration of compiler technology with database systems
is the most cost effective and versatile way forward. Thereby, a number of
specific features, which have been developed in compiler technology, can be
readily used by database systems. Most notably query optimizations have a
lot in common with optimizing compiler transformations.

Traditional query optimizers have long distinguished themselves from
optimizing compilers by relying on cost-based planning. A large search-space
of equivalent query plans for the execution of a given query is searched for
an optimal query plan. The optimality is determined through cost-based
metrics, where the cost of operations is calculated using statistics of the
database itself, most notably the size of the database relations. As the
contents and size of the relations in a database change, the optimal plan
for the execution of a query will change as well. Queries that have been
compiled once and do not adapt to changes in data will be outperformed by
the interpreted execution of query plans. Recompiling and optimizing a
query for each invocation of a query is too expensive an operation and the
gains might not weigh up.

In the last 30 years of research into optimizing compilers, the
consideration of run-time characteristics has been studied extensively. This
is caused by the fact that the transformation search space has become too
large to effectively search at compile time. Consequently, techniques such
as multi-version codes, Just-In-Time (JIT) compilation, iterative
compilation, run-time feedback mechanisms and trace scheduling were
developed. So, it is the potential of integrating query optimization with
already existing compiler optimizations taking run-time characteristics into
account that yields excellent opportunities for query optimizers to be
enhanced. In addition to common techniques used by query optimizers,
compilers are equipped with techniques that are not considered by query
optimizers such as enabling pre-fetching, instruction selection, code
compaction and automatic vectorization. As a result, compiler
transformations can outperform the cost-based planning that is performed by
traditional query optimizers.

The optimization methodology that is proposed in this paper is part of a
larger framework for the vertical integration of database
applications~\cite{rietveld-2015}.
Extensive vertical integration is not possible with traditional query
optimization techniques, because when code is generated from query
evaluation plans and combined with application code, further applicability
of compiler transformations is obscured. Therefore, it is important that
queries are transformed, optimized and combined with application code in a way
that compiler optimizations can still be successfully exploited.

We demonstrate in this paper that through the use of existing compiler
optimizations, a new generation of query optimizers can be built that are
capable of optimizing queries to a performance that is beyond that of
traditional query optimization (an average improvement of a factor of 3 and
in certain cases an improvement of up to a factor of 15).
An experimental evaluation is presented using the TPC-H
benchmark~\cite{tpc-h}. Note that the used compiler transformations are
the main query optimization techniques. This should not be confused with
other research into the use of compiler optimization in query optimization
such as~\cite{krikellas-2010,neumann-2011}, that propose compiler-based
techniques for the generation of efficient executable code from algebraic
query execution plans.

This paper is organized as follows. Section~\ref{sec:forelem-intro}
introduces the \emph{forelem} intermediate representation.
Section~\ref{sec:transformations} describes transformations that are defined
within the \emph{forelem} framework.
Section~\ref{sec:query-optimization} discusses how queries can be optimized
using the existing compiler transformations that are defined within the
\emph{forelem} framework and how the main techniques of query optimization
can be emulated using optimizing compiler techniques.
Section~\ref{sec:strategy} discusses strategies for the application of
transformations on \emph{forelem} loop nests and strategies for code generation.
Section~\ref{sec:experiments} compares the performance of
\emph{forelem}-optimized codes with that of a contemporary database
system using the queries from the TPC-H benchmark.
Section~\ref{sec:forelem-implementation} briefly discusses how the
\emph{forelem} framework is implemented.
Section~\ref{sec:related-work} discusses related work. Finally,
Section~\ref{sec:conclusions} lists our conclusions.

%% Section: Forelem introduction
\section{The Forelem Intermediate\\Representation}
\label{sec:forelem-intro}
In this section, the \emph{forelem} intermediate representation is
described, which forms an intermediate representation to which SQL queries are
transformed. Subsequently, the \emph{forelem} intermediate representation is
further optimized by ``compiler-type'' transformations.
The \emph{forelem} intermediate representation is centered around the
\emph{forelem} loop construct. Each \emph{forelem} loop iterates a
(multi)set of tuples. Tuples in these multisets are accessible with
subscripts, like ordinary arrays. The subscripts that are accessed through
an ``index set'' that is associated with the multiset.

Let us consider a simple query: \verb!SELECT A.a1! \verb!FROM A!
\verb!WHERE A.a2 = 7!.  A C code to evaluate this query could look as
follows:

\begin{alltt}
\textbf{for} (i = 0; i < len(A); i++)
  \textbf{if} (A[i].a2 == 7)
    add\_to\_result(A[i].a1)
\end{alltt}

\noindent
and in this code fragment, the \emph{for} loop iterates the full \verb!A! table,
the \emph{if}-statement selects matching tuples.

The main problem with this code fragment is that the looping structure is
explicit, which already gives a particular implementation of the query and
this limits the range of transformations and evaluation
choices~\cite{maier-1990}. It is apparent that a full iteration over
table \verb!A! is to be done, to check the value of \emph{a2} of each tuple.
This explicit looping structure excludes the possibility to, for example,
exploit an index on the \emph{a2} values. Additionally, more complex query
constructs, such as \emph{distinct} and \emph{group by}, require more
complicated code making it harder to apply transformations and hides the
actual problem at hand.

Ideally, only those rows are iterated for which the condition
$\mathit{a2} = 7$ holds true. This is similar to what an index on the
column \emph{a2} would accomplish. One way to accomplish this is to move the
definition of these conditions into the loop control structure, in our
case the \emph{for} statement. As a result, the explicit \emph{if}
statements are eliminated, which paves the way for the application of a
larger range of optimizations. The above query loop written using
\emph{forelem} looks like the following:

\begin{alltt}
\textbf{forelem} (i; i \(\in\) pA.a2[7])
  \(\mathscr{R} = \mathscr{R} \cup\) (A[i].a1)
\end{alltt}

\noindent
This code fragment is read as follows: with \verb!i!, iterate over each index
into table \verb!A! for which \verb!a2 == 7! holds true. For these $i$, we
append a tuple containing the value of \verb!a1! for index \verb!i! into
table \verb!A! to the result set $\mathscr{R}$.

Even though the \emph{forelem} loop appears to be very similar to a
\emph{foreach} loop that exists in many common programming languages, there
is one distinguishing feature. This concerns the notation
\verb!pA.a2[7]!. This denotes that a set of indices into table \verb!A!
will be returned for which the \verb!a2! field equals $7$. This is
similar to an \emph{index set} as is commonly used in DBMSs, and we will
also use this term to refer to the sets of indices we define here. The fact
that an index set contains indices is indicated by the prefix \emph{p}, from
pointer. Note that the order in which the indices appear in the index set is
not defined. From this follows that the exact semantics of how the table
\verb!A! will be iterated are not set in stone at this point.

Before proceeding, some further notation and terminology is introduced.
Let \verb!D! be a multiset representing a database table. \verb!D! can be
indexed with a subscript \verb!i! to get access to a tuple, or row, in
\verb!D!: \verb!D[i]!. A specific field of a row can be accessed with
\verb!D[i].field! where \verb!field! is a valid field of \verb!D!. Without
subscript, an entire column is selected resulting in a multiset containing
all values of that column: \verb!D.field!.

In \emph{forelem} codes that have been generated from SQL statements,
the loop body often outputs tuples to a temporary or result set. Temporary
sets are generally named $\mathscr{T}_1, \mathscr{T}_2, ..., \mathscr{T}_n$
and result sets (or output relations) $\mathscr{R}_1, \mathscr{R}_2, ...,
\mathscr{R}_n$. These temporary tables and result sets are both multisets.
The semantics that apply to multisets representing database tables apply to
temporary tables as well.

An \emph{index set} is a set containing subscripts $\mathtt{i} \in
\mathbb{N}$ into an array. Since each array subscript is typically processed
once per iteration of the array, these subscripts are stored in a regular
set. Index sets are named after the array they refer to, prefixed with
``p''.

\verb!pD! represents the index set of all subscripts into a database
table \verb!D!: $\forall \mathtt{t} \in \mathtt{D}: \exists!\;\mathtt{i} \in
\mathtt{pD}: \mathtt{D[i] = t}$. \verb!D! can also be a
temporary table $\mathscr{T}_n$. All rows of \verb!D! are visited if all members of
\verb!pD! have been used to subscript \verb!D!. Random access by subscript
into \verb!pD! is not possible, instead all accesses are done using the
$\in$ operator. $\mathtt{i} \in \mathtt{pD}$ stores the current index into
\verb!i! and advances \verb!pD! to the next entry in the index set.

Note that these index sets are not to be confused with indexes that are used
by database systems. As will become apparent, index sets are used to
control the iteration performed by the \emph{forelem} loop. The index sets
indicate \emph{what} subscripts to visit, but not in \emph{which} order as
\emph{forelem} loops do not impose a particular iteration order.

The part of the table that is selected using an index set can be narrowed
down by specifying conditions. For example, the index set denoted by
\verb!pD.field[k]! returns only those subscripts into \verb!D! for
which \verb!field! has value \verb!k!. This can also be expressed as follows:
\begin{equation*}
\mathtt{pD.field[k]} \equiv \{ \mathtt{i} \mid \mathtt{i} \in \mathtt{pD}
\wedge \mathtt{D[i].field == k}\}
\end{equation*}
\noindent
When a match on multiple fields is required, the single column name is
replaced with a tuple of column names:
\begin{multline*}
\mathtt{pD.(field1,field2)[(k\sb{1},k\sb{2})]} \equiv \\
\{\mathtt{i} \mid
\mathtt{i} \in \mathtt{pD} \wedge \mathtt{D[i].field1 == k\sb{1}} \wedge
\mathtt{D[i].field2 == k\sb{2}} \}
\end{multline*}
\noindent
Instead of a constant value, the values $\mathtt{k\sb{n}}$ can also be a
reference to a value from another table. To use such a reference, the table,
subscript into the table and field name must be specified, e.g.:
\verb!D[i].field!.

\subsection{Expressing Joins}
\label{sec:expressing-joins}
SQL statements with an arbitrary number of joins can be written as a
\emph{forelem} loop nest while preserving correctness of the results. This
is because both the SQL statement and the corresponding \emph{forelem} loop
nest set up the same Cartesian product. This fact will be used to reason
about the correctness of the translation and the conditions under which
transformations can be applied on the \emph{forelem} loops.
Consider the join query 
\verb!SELECT A.a1!
\verb!FROM A, C!
\verb!WHERE A.a_id! \verb!= C.a_id! \verb!AND! \verb!C.b_id = 13!
which is expressed in relational algebra as
$\pi_{\mathit{a1}}(\sigma_{A.a\_id = C.a\_id \wedge C.b\_id = 13}(A{\times}C))$,
or more commonly using the join operator:
$\pi_{\mathit{a1}}(\sigma_{C.b\_id = 13}(A\Join_{A.a\_id = C.a\_id}C))$.
Theoretically, a join is performed by first setting up the Cartesian product
over $A$ and $C$ and secondly selecting tuples which match the given
conditions. We can write the first relational algebra expression as a
\emph{forelem} loop nest. The part $A{\times}C$ can be written as follows:

\begin{alltt}
  \textbf{forelem} (i; i \(\in\) pA)
    \textbf{forelem} (j; j \(\in\) pC)
\(S\sb{1}\)   \(\mathscr{R} = \mathscr{R} \cup\) (A[i].*, C[j].*)
\end{alltt}

\noindent
where \verb!A[i].*! denotes all fields of table \verb!A! at subscript
\verb!i!. In the
result tuple all fields of \verb!A! are suffixed with $\mathtt{\sp{A}}$.
This loop nest sets up the Cartesian product $A{\times}C$ at statement
$S_1$, which stores the Cartesian product in $\mathscr{R}$. After executing
the loop nest, $\mathscr{R}$ is equivalent to what would be produced by the
relational algebra expression $A{\times}C$.

The selection operator $\sigma$ is implemented by making a pass over the
result table and only storing matching tuples in a new result table. Of the
matching tuples we only store the requested fields to implement the $\pi$
operator.

\begin{alltt}
  \textbf{forelem} (i; i \(\in\) p\(\mathscr{R}\))
    \textbf{if} (\(\mathscr{R}\)[i].a_id\(\sp{A}\) == \(\mathscr{R}\)[i].a_id\(\sp{C}\) &&
        \(\mathscr{R}\)[i].b_id\(\sp{C}\) == 13)
\(S\sb{1}\)    \(\mathscr{R}\sb{2} = \mathscr{R}\sb{2} \cup\) (\(\mathscr{R}\)[i].a1\(\sp{A}\))
\end{alltt}

\noindent
By application of a transformation that eliminates temporary tables from the
code, both loops can be merged into one:

\begin{alltt}
  \textbf{forelem} (i; i \(\in\) pA)
    \textbf{forelem} (j; j \(\in\) pC)
\(S\sb{1}\)    \textbf{if} (A[i].a_id == C[j].a_id && C[j].b_id == 13)
\(S\sb{2}\)      \(\mathscr{R} = \mathscr{R} \cup\) (A[i].a1)
\end{alltt}

\noindent
In general, we say that a perfectly nested \emph{forelem} loop nest of the
following form:

\begin{alltt}
 \textbf{forelem} (i\(\sb{1}\); i\(\sb{1} \in\) pT\(\sb{1}\))
   \textbf{forelem} (i\(\sb{2}\); i\(\sb{2} \in\) pT\(\sb{2}\))
     ...
       \textbf{forelem} (i\(\sb{n}\); i\(\sb{n} \in\) pT\(\sb{n}\))
\(S\sb{1}\)        \(\mathscr{R} = \mathscr{R} \cup\) (T\(\sb{1}\)[i\(\sb{1}\)].field, ..., T\(\sb{n}\)[i\(\sb{n}\)].field)
\end{alltt}

\noindent
sets up a Cartesian product of the tables $\mathtt{T}\sb{1},
\mathtt{T}\sb{2}, ..., \mathtt{T}\sb{n}$ at statement $S_1$. The Cartesian
product, or rather the part of the Cartesian product that is accessed, must
be preserved under any transformation for the query to yield correct
results.

\subsection{Expressing More Complicated Queries}
\label{sec:forelem-extensions}
Within the \emph{forelem} framework, a special syntax is available for
expressing the use of aggregate functions. An aggregate function typically
has three stages: initialization, update and finalization. These stages serve
to initialize any variables, update the variables for each tuple that is
processed and to come to a final result. To use aggregate functions in
the \emph{forelem} framework, the following functions are supplied that
represent the different stages of the aggregate function: \verb!agg_init!
for initialization, \verb!agg_step! to update the variables for a tuple,
\verb!agg_finish! to finish the computation of the aggregate. Additionally,
\verb!agg_result! returns the final result of the aggregate. The first
argument to all these functions is the handle number, such that multiple
aggregates can be computed at the same time.

As an example consider the query \verb!SELECT SUM(A.a1) FROM A!. This
query is written in the \emph{forelem} intermediate as follows:

\small
\begin{alltt}
agg_init(0, SUM);
\textbf{forelem} (i; i \(\in\) pA)
  agg_step(0, A[i].a1);
agg_finish(0);
\(\mathscr{R} = \mathscr{R} \cup\) (agg_result(0))
\end{alltt}
\normalsize

Using the aggregate syntax group by queries can be easily transformed. As an
example, consider the following SQL query:
\verb!SELECT A.field2,! \verb!MIN(A.field3)!
\verb!FROM A!
\verb!GROUP BY A.field2!
This query can be transformed in the \emph{forelem} intermediate
representation as follows and will be referred to as code sample (1):

\begin{alltt}
\textbf{forelem} (i; i \(\in\) pA)
  \(\mathscr{T} = \mathscr{T} \cup\) (A[i].field2, A[i].field3)
\textbf{forelem} (i; i \(\in\) p\(\mathscr{T}\))
  \(\mathscr{G} = \mathscr{G} \cup\) (\(\mathscr{T}\)[i].field2)
distinct(\(\mathscr{G}\))
\textbf{forelem} (i; i \(\in\) p\(\mathscr{G}\))
\{
  agg_init(0, MIN);
  \textbf{forelem} (j; j \(\in\) p\(\mathscr{T}\).field2[\(\mathscr{G}\)[i].field2])
    agg_step(0, \(\mathscr{T}\)[j].field3);
  agg_finish(0);
  \(\mathscr{R} = \mathscr{R} \cup\) (\(\mathscr{G}\)[i].field2, agg_result(0))
\}
\end{alltt}

\noindent
Note the use of $\mathtt{distinct}(\mathscr{G})$ which denotes that
duplicate values from $\mathscr{G}$ are eliminated. The \emph{forelem} loop
above is an initial expression of the query in \emph{forelem}, which will be
subsequently subjected to transformations at the \emph{forelem} level.
Subqueries are handled by initially expressing these as functions, which are
called from the main loop nest. For example, consider:

\begin{alltt}
\textbf{SELECT} A.a1
\textbf{FROM} A, B, C
\textbf{WHERE} A.a_id = C.a_id \textbf{AND} C.b_id = B.b_id
\textbf{AND} B.b2 = "str1"
\textbf{AND} A.a_id \textbf{IN} (\textbf{SELECT} A2.a_id
              \textbf{FROM} A A2, B B2, C C2
              \textbf{WHERE} A2.a_id = C2.a_id
              \textbf{AND} C2.b_id = B2.b_id
              \textbf{AND} B2.b2 = "str2")
\end{alltt}

\noindent
Then this query is translated to the following \emph{forelem} loops:

\begin{alltt}
\textbf{function} subquery ()
\{
  \(\mathscr{T} = \emptyset\)
  \textbf{forelem} (i; i \(\in\) pB.b2["str2"])
    \textbf{forelem} (j; j \(\in\) pC.b_id[B[i].b_id])
      \textbf{forelem} (k; k \(\in\) pA.a_id[C[j].a_id])
        \(\mathscr{T} = \mathscr{T} \cup\) A[k].a_id
  \textbf{return} \(\mathscr{T}\)
\}

\textbf{forelem} (i; i \(\in\) pB.b2["str1"])
  \textbf{forelem} (j; j \(\in\) pC.b_id[B[i].b_id])
    \textbf{forelem} (k; k \(\in\) pA.a_id[C[j].a_id])
      \textbf{if} (A[k].a_id \(\in \)subquery())
        \(\mathscr{R} = \mathscr{R} \cup\) (A[k].a1)
\end{alltt}

\noindent
Note that function \verb!subquery! does not have any arguments. Correlated
subqueries will have function arguments. Similar to the translation of group
by queries, this is an initial expression of the query which will be
subjected to transformations in order to optimize the query.

\section{Transformations Within The Forelem Intermediate Representation}
\label{sec:transformations}
Due to the nature of \emph{forelem} as a simple loop, many existing loop
transformations can be applied on these loops. Existing loop transformations
that can be readily re-targeted to \emph{forelem} loops include Loop
Interchange, Loop Fusion~\cite{kennedy-1994} and Loop Invariant Code Motion.
Other common compiler techniques that are used within the \emph{forelem}
framework are (Function) Inline and Dead Code Elimination. Dead Code
Elimination has for instance been adapted to be able to detect unused
(temporary) tables and to remove \emph{forelem} loops that generate such
tables. Similarly, common analysis techniques
are re-used. Firstly, data dependence
analysis~\cite{kuck-1981,allen-phd,allen-1987,zima-1991} is used to
determine whether a given transformation can be applied without affecting
the correctness of a program. Secondly, def-use
analysis~\cite{allen-1976,kennedy-1981} is used to find unused variables, or
to infer the current value of a variable by looking at preceding definitions
of the variable in the def-use chain.

Because these transformations and analysis techniques are well established
within the literature and are readily re-targeted, we will not further
discuss these transformations in detail in this section. Rather, we will
introduce new transformations that have been defined within the
\emph{forelem} framework especially for the purpose of query optimization.
Within the \emph{forelem} framework queries are optimized by the application
of such transformations on \emph{forelem} loop nests, contrary to
traditional query planning where transformations are carried out on a
relational algebra tree.

\subsection{Iteration Space Expansion}
\label{sec:iteration-space-expansion}
Scalar Expansion is a transformation that is typically used to enable
parallelization of loop nests. Consider the following loop:

\begin{alltt}
\textbf{for} (k = 1; k <= N; k++)
  \{
    tmp = A[k] + B[k];
    C[k] = tmp / 2;
  \}
\end{alltt}

\noindent
Due to the loop-carried anti-dependence of \verb!tmp!, subsequent iterations
cannot write to \verb!tmp! before \verb!tmp! has been used in the assignment
to \verb!C[k]!. This is solved by the Scalar Expansion transformation which
expands the scalar \verb!tmp! to a vector:

\begin{alltt}
\textbf{for} (k = 1; k <= N; k++)
  \{
    tmp[k] = A[k] + B[k];
    C[k] = tmp[k] / 2;
  \}
\end{alltt}

\noindent
Now that the loop-carried dependence has been broken, the loop can be
parallelized.

Within the \emph{forelem} framework a transformation known as Iteration
Space Expansion is defined. This transformation expands the iteration space
of a \emph{forelem} loop by removing conditions on its index set.
For a loop of the form, with \verb!SEQ! denoting a sequence of statements:

\begin{alltt}
\textbf{forelem} (i; i \(\in\) pA.field[X])
  SEQ;
\end{alltt}

\noindent
the following steps are performed:

\begin{enumerate}
\item the condition \verb!A[i].field == X! is removed, which expands the
iteration space so that the entire array \verb!A! is visited,
\item scalar expansion is applied on all variables that are written to in
the loop body denoted by \emph{SEQ} and references to these variables are
subscripted with the value tested in the condition, in this case
\verb!A[i].field!,
\item all references to the scalar expanded variables after the loop are
rewritten to reference subscript \verb!X! of the scalar expanded variable.
\end{enumerate}

\subsection{Table Propagation}
The Table Propagation transformation is similar to Scalar Propagation that
is typically performed by compilers. In Scalar Propagation, the use of
variables whose value is known at compile-time is substituted with that
value.  In Table Propagation, the use of a temporary table of which the
contents are known is replaced with a loop nest that generates the same
contents as the temporary table. This eliminates unnecessary copying of data
to create the temporary table, but also enables further transformations
because the loop nest that generates the contents of the temporary table can
now be considered together with the loop nest that iterates the temporary
table.  For example, consider the following \emph{forelem} loops:

\begin{alltt}
\textbf{forelem} (i; i \(\in\) pX.field2[value])
  \(\mathscr{T} = \mathscr{T} \cup\) (X[i].field3)
\textbf{forelem} (i; i \(\in\) p\(\mathscr{T}\))
  \(\mathscr{R} = \mathscr{R} \cup\) (\(\mathscr{T}\)[i].field3)
\end{alltt}

\noindent
The first loop generates a table $\mathscr{T}$, which is iterated by the
second loop. The table $\mathscr{T}$ is being ``streamed'' between these
consecutive loops. Table $\mathscr{T}$ can be propagated to the second loop
nest:

\begin{alltt}
\textbf{forelem} (i; i \(\in\) pX.field2[value])
  \(\mathscr{T} = \mathscr{T} \cup\) (X[i].field3)
\textbf{forelem} (i; i \(\in\) pX.field2[value])
  \(\mathscr{R} = \mathscr{R} \cup\) (X[i].field3)
\end{alltt}

\noindent
After application of this transformation, temporary tables that are
generated (by the first loop in this example) are often left unused and can
be eliminated by application of the Dead Code Elimination transformation.

\subsection{Materialization \& Concretization}
\label{sec:materialization}
\emph{forelem} loops only specify how a set of tuples is iterated but do not
specify how these tuples are stored. When C/C++ code is generated from
\emph{forelem} loops, it must be known how the tuples are to be stored,
otherwise no code can be generated. Within the \emph{forelem} framework
Materialization and Concretization transformations are
defined~\cite{rietveld-2013-cpc}, which are used to devise a data storage
format for each collection of tuples. Using these transformations, many
different storage methods for tuples can be generated automatically. This
includes row-wise and column-wise storage orders. Conversion between these
two layouts is a matter of a trivial transformation within the
\emph{forelem} framework.

An in-depth discussion of how these techniques work and interact with other
transformations that are defined is outside the scope of this paper. Within
the context of this paper, we assume that a table \verb!Table! has a
materialized sibling \verb!PTable!, which stores the content of the tables
in a certain format and order on the disk.

Next to determining how the tables are stored, the Materialization
techniques are also used to determine a storage format (or in fact data
structure) for the index sets. Index sets that are used within a
\emph{forelem} loop nest are either translated to an \emph{if}-statement or
an index set is generated at run-time. This difference can be illustrated as
follows. Consider the \emph{forelem} loop:

\begin{alltt}
\textbf{forelem} (i; i \(\in\) pTable.field[value])
  SEQ using T[i];
\end{alltt}

\noindent
In the former case, an index set is not generated and the result is:

\begin{alltt}
\textbf{forelem} (i; i \(\in \mathbb{N}\sb{Table}\))
  \textbf{if} (PTable[i].field == value)
    SEQ using PTable[i];
\end{alltt}

\noindent
where $\mathbb{N}_{Table}$ denotes the number of tuples in \verb!Table!. If
this \emph{forelem} loop is executed many times for different instances of
\verb!value! this is not efficient as for every execution the full table
\verb!Table! has to be processed. The latter case alleviates this by
generating an index set such that \verb!Table! is only iterated once.
We say that an index set is materialized (Index Set Materialization) to
result in:

\begin{alltt}
\textbf{forelem} (i; i \(\in \mathbb{N}\sb{Table}\))
  pPTable.field[PTable[i].field] =
         pPTable.field[PTable[i].field] \(\cup\) (i)

\textbf{forelem} (i; i \(\in\) pPTable.field[value])
  SEQ using PTable[i];
\end{alltt}

\noindent
In this code fragment, \verb!pPTable! is an associative array that relates
values \verb!PTable[i].field! to subscripts \verb!i! into \verb!PTable!.
Subsequently, this index set is used to access the table stored on disk.

The actual generation of the index set is similar to a compiler technique
known as data copying~\cite{lam-1991,temam-1993}. In data copying, a partial
copy is made of a block of data that is processed by a loop. Although there
is a cost involved in making this copy, the copy results in a much better
utilization of the cache and thus in a significant increase in performance.
The initial cost of making the copy is redeemed. In the above example, a
partial copy has been made of \verb!PTable! by copying values of
\verb!field! and corresponding subscript values into the table on disk into
a temporary associative array that represents the index set. Instead of
storing subscript values, it is also possible to store tuples of values, for
instance when it is known that only 1 or 2 fields of \verb!PTable! would be
accessed.

Now that the content of the index set has been established, it needs to be
determined \emph{how} to store this content. This phase is known as
Concretization. During this phase, the associative array \verb!pPTable! is
concretized to a suitable data structure. Prime examples are for example a
balanced tree or hash table, such as implemented in for instance the Boost
libraries, or direct access through an array.  From the definition of this
storage format (selection of the data structure), code can be generated that
instantiates the index set at run-time and to access this index set.

\subsection{Index Set Pruning}
Once index sets have been materialized using the process described in the
previous subsection, further transformations can be defined that affect the
contents of the index set. One such transformation is Index Set Pruning.
With Index Set Pruning, the contents of an index set are reduced using
additional conditions that exist on tuples iterated by this index set. As a
result, less memory is required to store the index set and the generation of
the index set is less computationally expensive.  As an example, consider:

\begin{alltt}
\textbf{forelem} (i; i \(\in \mathbb{N}\sb{Table}\))
  pPTable.field[PTable[i].field] =
         pPTable.field[PTable[i].field] \(\cup\) (i)
\end{alltt}
\begin{alltt}
\textbf{forelem} (i; i \(\in\) pPTable.field[value1])
  \textbf{if} (PTable[i].field == value2)
    SEQ;
\end{alltt}

\noindent
\verb!SEQ! will not be executed for tuples that do not satisfy the condition
posed in the \emph{if}-statement. If the materialized index set
\verb!pPTable.field! is not used in a different context for which different
additional conditions might hold, it is not necessary to store entries in
the index set for tuples that do not adhere to this condition. The index
set can thus be pruned, by moving the \emph{if}-statement to the loop that
generates the materialized index set:

\begin{alltt}
\textbf{forelem} (i; i \(\in \mathbb{N}\sb{Table}\))
  \textbf{if} (PTable[i].field == value2)
    pPTable.field[PTable[i].field] =
           pPTable.field[PTable[i].field] \(\cup\) (i)

\textbf{forelem} (i; i \(\in\) pPTable.field[value1])
  SEQ;
\end{alltt}

\noindent
This transformation is useful in the following example:

\begin{alltt}
\textbf{forelem} (i; i \(\in\) pTable1.(f2,f3)[(v1, v2)])
  \textbf{forelem} (j; j \(\in\) pTable2.f1[Table1[i].f1])
    \textbf{forelem} (k; k \(\in\) pTable3.f1[Table2[j].f2])
      \textbf{forelem} (l; l \(\in\) pTable4.f1[Table3[k].f1])
        \textbf{if} (Table4[l].f2 == v3)
          SEQ;
\end{alltt}

\noindent
In this example, the transformation engine decided to move the tests of the
conditions on \verb!Table1! to the outermost loop, because two conditions
are tested and potentially prunes the search space by a larger extent. Due
to the dependences between the other tables, the condition for \verb!Table4!
is only tested in the inner loop. First, the index set iterated by the
innermost loop is materialized to result in:

\begin{alltt}
\textbf{forelem} (i; i \(\in\) pTable1.(f2,f3)[(v1, v2)])
  \textbf{forelem} (j; j \(\in\) pTable2.f1[Table1[i].f1])
    \textbf{forelem} (k; k \(\in\) pTable3.f1[Table2[j].f2])
      \textbf{forelem} (l; l \(\in\) pPTable4.f1[Table3[k].f1])
        \textbf{if} (PTable4[l].f2 == v3)
          SEQ;
\end{alltt}

\noindent
As a next step, the \emph{if}-statement can be eliminated by Index Set
Pruning:

\begin{alltt}
\textbf{forelem} (i; i \(\in\) pTable1.(f2,f3)[(v1, v2)])
  \textbf{forelem} (j; j \(\in\) pTable2.f1[Table1[i].f1])
    \textbf{forelem} (k; k \(\in\) pTable3.f1[Table2[j].f2])
      \textbf{forelem} (l; l \(\in\) pPTable4.f1[Table3[k].f1])
        SEQ;
\end{alltt}

\noindent
The loop generating \verb!pPTable4.field! has become:

\begin{alltt}
\textbf{forelem} (i; i \(\in \mathbb{N}\sb{Table4}\))
  \textbf{if} (PTable4[i].f2 == v3)
    pPTable4.f2[PTable4[i].f2] =
            pPTable4.f2[PTable4[i].f2] \(\cup\) (i)
\end{alltt}

\noindent
We will now describe another transformation, Index Set Combination, which
leads to a further improvement of the performance of this example.

\subsection{Index Set Combination}
When an index set is solely used to test for a certain property and not used
to access the corresponding table, there is a good chance this index set can
be combined with another index set. Observe the loop iterating \verb!Table4!
in the example of the previous subsection. Assume that the iterator variable
\verb!l! is not used in \verb!SEQ!, so \verb!Table4! is not accessed within
the loop body. We can then combine this index set with the index set for
\verb!Table3! (materialized to \verb!pPTable3.field1!, so that only
subscripts \verb!k! are iterated that also satisfy the conditions
tested by the index set for \verb!Table4!). This combination process is
referred to as Index Set Combination.

Consider that the loops generating the index sets for \verb!Table3! and
\verb!Table4! are:

\begin{alltt}
\textbf{forelem} (i; i \(\in \mathbb{N}\sb{Table3}\))
  pPTable3.f1[PTable3[i].f1] =
          pPTable3.f1[PTable3[i].f1] \(\cup\) (i)

\textbf{forelem} (i; i \(\in \mathbb{N}\sb{Table4}\))
  \textbf{if} (PTable4[i].f2 == v3)
    pPTable4.f2[PTable4[i].f2] =
            pPTable4.f2[PTable4[i].f2] \(\cup\) (i)
\end{alltt}

\noindent
The condition on \verb!Table4! can be combined into these for \verb!Table3!
as follows:

\begin{alltt}
\textbf{forelem} (i; i \(\in \mathbb{N}\sb{Table4}\))
  \textbf{if} (PTable4[i].f2 == v3)
    pPTable4.f2[PTable4[i].f2] =
            pPTable4.f2[PTable4[i].f2] \(\cup\) (i)

\textbf{forelem} (i; i \(\in \mathbb{N}\sb{Table3}\))
  \textbf{if} (is_not_empty(pPT4.f1[PTable3[k].f1]))
    pPTable3.f1[PTable3[i].f1] =
            pPTable3.f1[PTable3[i].f1] \(\cup\) (i)
\end{alltt}

\noindent
The combination of the index sets is performed based on the condition
\verb!Table4.field1 == Table3.field1! that is encoded in the original index
sets. This condition is used to set up the ``is not empty'' test. As a
consequence, \verb!Table4! no longer needs to be iterated in the main
loop nest, resulting in:

\newpage
\begin{alltt}
\textbf{forelem} (i; i \(\in\) pTable1.(f2,f3)[(v2, v3)])
  \textbf{forelem} (j; j \(\in\) pTable2.f1[Table1[i].f1])
    \textbf{forelem} (k; k \(\in\) pPTable3.f1[Table2[j].f2])
      SEQ;
\end{alltt}

\noindent
Note that a further transformation is possible on the loops that generate
the index sets. The \emph{if}-statement checking the \verb!is_not_empty!
condition can be substituted with the loop actually generating the checked
index set (Table Propagation). Subsequently, the loop generating
\verb!pPTable4! can be eliminated at the cost of iterating table
\verb!PTable4! multiple times. In some cases, this is beneficial, for
instance when \verb!PTable4! is small and fits CPU cache.

%% Section: compiler transformations
\section{Performing Query Optimization With Compiler Transformations}
\label{sec:query-optimization}
In this section, we discuss the optimization of queries through the use of
existing compiler transformations, rather than the use of query planning
techniques used by traditional query optimizers. In particular, we
demonstrate that the main techniques of query optimization can be emulated
using optimizing compiler techniques. The main techniques that will be
discussed have been distilled from~\cite{ramakrishnan-2003,chaudhuri-1998}
and include operations on the algebraic query tree and different algorithms
for the execution of relational operators.

First two optimizations on a relational query tree, Join Reordering and
Selection Pushing are considered. After this kind of transformations has
been carried out and an order in which the joins are to be processed has
been selected, a query optimizer needs to select for each join operator in
the query tree with which algorithm to evaluate that join. Common algorithms
are Nested Loops, Block Nested Loops, Index Nested Loops and Hash Join. So,
secondly, the correspondences between the \emph{forelem} framework and these
algorithms will be described. Within the \emph{forelem} framework, no fixed
implementations of such join operators are implemented. Therefore,
implementations of these operators are described through the application of
transformations within the \emph{forelem} framework.

\subsection{Join Reordering}
The most important task of a traditional query optimizer is to determine in
which order the joins in a query should be processed.
To determine such an order, a query optimizer considers the
search space consisting of equivalent query plans and for each plan computes
the cost of executing this plan. This cost depends on the number of disk
I/Os that have to be performed and the estimated cost of executing every
relational operator that is present in the query tree.

As has been discussed in Section~\ref{sec:expressing-joins}, in the
\emph{forelem} framework a query performing multiple joins is represented as
a nested loop. At each nesting level a different table is accessed. Using
the Loop Interchange transformation, the order of loops in the loop nest is
reordered. Although the order of the loops is changed, the cross product
that is generated in the loop body is still the same. So, the result of the
query is not affected. This is essentially the same operation as join
reordering. When joins are reordered in the query plan tree, the result of
the query is not affected either.

Loop Interchange sets up a search space of all possible orderings of the
loops in a loop nest. A compiler can select an appropriate order at
compile-time through the use of heuristics, such as putting loops imposing
most conditions as the outermost loops, and by incorporating run-time
information in an iterative compilation~\cite{knijnenburg-2003,fursin-2005}
process. By exploiting the collected run-time information, the compiler will
select better performing loop orders each time the code is executed and the
compiler is also enabled to adapt to changes in the data set.

\subsection{Selection Pushing}
Next to join, \emph{selection} is another important relational operator.
Consider a relational
query tree which contains a join and selection operator for the same table.
If first the join is performed and then selection, then at the selection
operator many rows are discarded that do not satisfy the condition. So, a
lot of excess work has been performed. To avoid this, query optimizers apply
Selection Pushing to push selection operators to be performed before the
join operator.

Within the \emph{forelem} framework Selection Pushing is implemented using
the Loop Invariant Code Motion operator. To see, consider the following
example:

\begin{alltt}
  \textbf{forelem} (i; i \(\in\) pA)
    \textbf{forelem} (j; j \(\in\) pC)
      \textbf{if} (A[i].field < 20)
\(S\sb{1}\)   \(\mathscr{R} = \mathscr{R} \cup\) (A[i].*, C[j].*)
\end{alltt}

\noindent
The if-statement is executed for all rows that are the result of the cross
product between \verb!A! and \verb!C!. The statement $S_1$ is only executed
for rows of the cross product that satisfy the condition
\verb!A[i].field < 20!. Observe that the value \verb!A[i].field! is constant
under the loop iterated by \verb!j!. That is, if \verb!j! is incremented,
the value \verb!A[i].field! will not change. The if-statement is invariant
under the loop iterated by \verb!j! and can be moved out of this loop.

\begin{alltt}
  \textbf{forelem} (i; i \(\in\) pA)
    \textbf{if} (A[i].field < 20)
      \textbf{forelem} (j; j \(\in\) pC)
\(S\sb{1}\)   \(\mathscr{R} = \mathscr{R} \cup\) (A[i].*, C[j].*)
\end{alltt}

\noindent
Rows in \verb!A! that do not satisfy the condition are now not considered
when the cross product is created. Important is that statement $S_1$ is
still executed for rows of the cross product that satisfy the condition
\verb!A[i].field < 20!. The actual outcome of the query has not changed,
since statement $S_1$ is still executed for the same set of the rows, but
the execution time has been improved because the loop iterated by \verb!j!
is only executed for qualifying rows of \verb!A! instead of all rows of
\verb!A!. Similar to Selection Pushing, the selection operator (the
if-statement) has been moved to before the cross product is created (before
the join).

\subsection{Nested Loops Join}
The first implementation of the join operator that is considered is Nested
Loops Join. To describe the different implementations a simple example will
be used:
\verb!SELECT A.a!
\verb!FROM A, B!
\verb!WHERE A.b == B.b;!
which corresponds to the following \emph{forelem} loop nest:

\newpage
\begin{alltt}
\textbf{forelem} (i; i \(\in\) pA)
  \textbf{forelem} (j; j \(\in\) pB.b[A[i].b])
    \(\mathscr{R} = \mathscr{R} \cup\) (A[i].a)
\end{alltt}

\noindent
In Nested Loops Join, a nested loop is used to generate the cross product.
The conditions are tested within the loop body. This roughly corresponds to
the following \emph{forelem} loop nest wherein no index set is used:

\begin{alltt}
\textbf{forelem} (i; i \(\in\) pA)
  \textbf{forelem} (j; j \(\in\) pB)
    \textbf{if} (A[i].b == B[j].b)
      \(\mathscr{R} = \mathscr{R} \cup\) (A[i].a)
\end{alltt}

\noindent
Tables \verb!A! and \verb!B! are unordered collections of tuples. A
database table that is stored on disk does have an order. With a subscript
to a tuple the address of the tuple on disk can be determined. So, we
consider database tables as stored on disk to be the materialized tables
\verb!PA!, \verb!PB!. Through the application of Loop-Independent
Materialization on both of the loops in the loop nest, the actual
implementation of Nested Loops Join is found:

\begin{alltt}
\textbf{forelem} (i; i \(\in \mathbb{N}\sb{A}\))
  \textbf{forelem} (j; j \(\in \mathbb{N}\sb{B}\))
    \textbf{if} (PA[i].b == PB[j].b)
      \(\mathscr{R} = \mathscr{R} \cup\) (PA[i].a)
\end{alltt}

\noindent
In this code fragment, both tables are iterated entirely and the condition
for the rows is tested within the loop body.

\subsection{Block Nested Loops Join}
\label{sec:blocked-nested-loops-join}
Block Nested Loops Join follows analogously from Nested Loops Join if Loop
Blocking is incorporated in the transformation chain. The table that is
iterated by the outermost loop is blocked. It is important that such a block
fits in main memory, while the inner loop iterates the second table.

As the first step, table \verb!A! is partitioned into blocks. This is done
by partitioning the index set:
$\mathtt{pA} = \mathtt{p}_{1}\mathtt{A} \cup \mathtt{p}_{2}\mathtt{A} \cup
\ldots \cup \mathtt{p}_{N}\mathtt{A}.$
On the initial \emph{forelem} loop nest the Loop Blocking transformation is
applied that utilizes this partitioning of \verb!A!:

\begin{alltt}
\textbf{forelem} (l; l \(\in\) N)
  \textbf{forelem} (i; i \(\in\) p\(\sb{l}\)A)
    \textbf{forelem} (j; j \(\in\) pB.b[A[i].b])
      \(\mathscr{R} = \mathscr{R} \cup\) (A[i].a)
\end{alltt}

\noindent
On this loop, materialization can be applied likewise to how this was done
in the previous subsection:

\begin{alltt}
\textbf{forelem} (l; l \(\in\) N)
  \textbf{forelem} (i; i \(\in \mathbb{N}\sb{l{\sb{A}}}\))
    \textbf{forelem} (j; j \(\in \mathbb{N}\sb{B}\))
      \textbf{if} (PA[i].b == PB[j].b)
        \(\mathscr{R} = \mathscr{R} \cup\) (PA[i].a)
\end{alltt}

This particular join algorithm is often optimized by creating a hash table
for each block of \verb!A!. This implies that can index set
$\mathtt{p}\sb{l}\mathtt{A.b}$ is generated for each block of \verb!A!. For
every row that is read from \verb!B! this index set is queried to determine
whether there is a corresponding row in \verb!A!. So, first Loop Interchange
is applied to result in:

\begin{alltt}
\textbf{forelem} (l; l \(\in\) N)
  \textbf{forelem} (j; j \(\in\) pB)
    \textbf{forelem} (i; i \(\in\) p\(\sb{l}\)A.b[B[j].b])
      \(\mathscr{R} = \mathscr{R} \cup\) (A[i].a)
\end{alltt}

\noindent
As a second step, the tables \verb!A! and \verb!B! need to be materialized.
In conjunction with this, the index set also must be materialized so that it
contains subscripts to a materialized table instead of a non-materialized
table. The result of materializing the tables as well as the index set is:

\begin{alltt}
\textbf{for} (l; l \(\in\) N)
\{
  \textbf{forelem} (j; j \(\in \mathbb{N}\sb{l\sb{A}}\))
    p\(\sb{l}\)PA.b[PA[j].b] = j;
  \textbf{forelem} (j; j \(\in \mathbb{N}\sb{B}\))
    \textbf{forelem} (i; i \(\in\) p\(\sb{l}\)PA.b[PB[j].b])
      \(\mathscr{R} = \mathscr{R} \cup\) (PA[i].a)
\}
\end{alltt}

\subsection{Index Nested Loops Join}
Index Nested Loops Join takes advantage of indexes on tables that are
already present. Such indexes are created explicitly in the database system
and are kept up to date as rows are added, removed and updates in tables.
Consider again our starting point:

\begin{alltt}
\textbf{forelem} (i; i \(\in\) pA)
  \textbf{forelem} (j; j \(\in\) pB.b[A[i].b])
    \(\mathscr{R} = \mathscr{R} \cup\) (A[i].a)
\end{alltt}

\noindent
Assume that an index set \verb!pA.b! is available in the system. Then
transformations must be carried out such that advantage can be taken from
this index set. In this case, the Loop Interchange transformation is carried
out:
\begin{alltt}
\textbf{forelem} (j; j \(\in\) pB)
  \textbf{forelem} (i; i \(\in\) pA.b[B[j].b])
    \(\mathscr{R} = \mathscr{R} \cup\) (A[i].a)
\end{alltt}
\noindent
In the resulting loop nest, the index set \verb!pA.b! is indeed used. During
materialization, \verb!pA.b! is lowered to the existing index on table
\verb!A!, similar to how \verb!A! is lowered to the existing array
\verb!PA!.

\subsection{Hash Join}
The Hash Join algorithm consists of two phases: (1) both tables are
partitioned on the join attribute, (2) for each partition process the
partitions of both relations to produce the join results.
The first phase is implemented within the \emph{forelem} framework by the
application of Loop Blocking. As we have seen above, with Loop Blocking a
table is partitioned. In this particular case, both tables are partitioned
with the additional requirement that corresponding partitions must contain
all rows that are to be joined with one another. So, the partitioning must
be join attribute aware. Let
$\mathtt{A} = \mathtt{A}_{1} \cup \mathtt{A}_{2} \cup \ldots \cup
\mathtt{A}_{N}$ and
$\mathtt{B} = \mathtt{B}_{1} \cup \mathtt{B}_{2} \cup \ldots \cup
\mathtt{B}_{N}$.
The resulting code for the hash join algorithm is:
\vskip 0.2cm
\begin{alltt}
\textbf{forelem} (l; l \(\in\) N)
  \textbf{forelem} (i; i \(\in\) pA\(\sb{l}\))
    \textbf{forelem} (j; j \(\in\) pB\(\sb{l}\).b[A\(\sb{l}\)[i].b])
      \(\mathscr{R} = \mathscr{R} \cup\) (A\(\sb{l}\)[i].a)
\end{alltt}
\noindent
As a next step, the loop nest is materialized. This includes materializing
the index set has we have have treated the creation of an in-memory hash
already in Section~\ref{sec:blocked-nested-loops-join} above. This results
in the following code fragment:
\begin{alltt}
\textbf{for} (l; l \(\in\) N)
\{
  \textbf{forelem} (j; j \(\in \mathbb{N}\sb{A\sb{l}}\))
    pPA\(\sb{l}\).b[PA\(\sb{l}\)[j].b] = j;
  \textbf{forelem} (j; j \(\in \mathbb{N}\sb{B\sb{l}}\))
    \textbf{forelem} (i; i \(\in\) pPA\(\sb{l}\).b[PB\(\sb{l}\)[j].b])
      \(\mathscr{R} = \mathscr{R} \cup\) (PA\(\sb{l}\)[i].a)
\}
\end{alltt}

\subsection{Sorted Aggregation}
In the Sorted Aggregation strategy, a group-by query is evaluated by
performing a single pass through a sorted table. A materialized code that
applies the sorted aggregation strategy is derived from an aggregate query
transformed to \emph{forelem} (the starting point, see
Section~\ref{sec:forelem-extensions} for an example) using techniques to
materialize (temporary) tables to sorted arrays. Subsequently, index sets
with a condition that tests for equality on sorted arrays can be
materialized without creating an associative array.  Rather, these index
sets can be evaluated using the fact that the array that is accessed is
sorted, the array is iterated until a row is encountered of which the value
tested by the index set condition is larger than the value that is part of
the condition.

Once these index sets have been materialized in this fashion, further
transformations can be carried out using the properties of the tables that
are accessed. Since $\mathscr{T}$ and $\mathscr{G}$ are temporary tables, it
is known how these are derived from the def-use chains. Using this
knowledge, transformations can be carried out to eliminate redundant loop
iterations (making use of the property that the tables are sorted and a
value can only occur in $\mathscr{G}$ once) and a different form of Table
Propagation can be applied to eliminate usage of $\mathscr{G}$ leading to
the query being evaluated using a single pass of $\mathscr{T}$.
The result is a code that evaluates the group-by query using the sorted
aggregation strategy.

\subsection{Hash Aggregation}
The Hash Aggregation strategy can be obtained by applying the Iteration
Space Expansion transformation, discussed in
Section~\ref{sec:iteration-space-expansion}. Consider \emph{forelem} code
sample (1) described in Section~\ref{sec:forelem-extensions}. After inling
the aggregate functions and performing Iteration Space Expansion this loop
becomes:

\begin{alltt}
\textbf{forelem} (i; i \(\in\) pA)
  \(\mathscr{T} = \mathscr{T} \cup\) (A[i].field2, A[i].field3)
\textbf{forelem} (i; i \(\in\) p\(\mathscr{T}\))
  \(\mathscr{G} = \mathscr{G} \cup\) (\(\mathscr{T}\)[i].field2)
distinct(\(\mathscr{G}\))
\textbf{forelem} (i; i \(\in\) p\(\mathscr{G}\))
\{
  tmp[] = INT_MAX;
  \textbf{forelem} (j; j \(\in\) p\(\mathscr{T}\))
  \textbf{if} (\(\mathscr{T}\)[j].field3 < tmp[\(\mathscr{T}\)[j].field2])
      tmp[\(\mathscr{T}\)[j].field2] = \(\mathscr{T}\)[j].field3;
  \(\mathscr{R} = \mathscr{R} \cup\) (\(\mathscr{G}\)[i].field2, tmp[\(\mathscr{G}\)[i].field2])
\}
\end{alltt}

\noindent
As a result, the inner \emph{forelem} loop is now invariant under the
iterator variable \verb!i! and can be moved out of the loop:

\begin{alltt}
\textbf{forelem} (i; i \(\in\) pA)
  \(\mathscr{T} = \mathscr{T} \cup\) (A[i].field2, A[i].field3)
\textbf{forelem} (i; i \(\in\) p\(\mathscr{T}\))
  \(\mathscr{G} = \mathscr{G} \cup\) (\(\mathscr{T}\)[i].field2)
distinct(\(\mathscr{G}\))
tmp[] = INT_MAX;
\textbf{forelem} (j; j \(\in\) p\(\mathscr{T}\))
  \textbf{if} (\(\mathscr{T}\)[j].field3 < tmp[\(\mathscr{T}\)[j].field2])
    tmp[\(\mathscr{T}\)[j].field2] = \(\mathscr{T}\)[j].field3;
\textbf{forelem} (i; i \(\in\) p\(\mathscr{G}\))
  \(\mathscr{R} = \mathscr{R} \cup\) (\(\mathscr{G}\)[i].field2, tmp[\(\mathscr{G}\)[i].field2])
\end{alltt}

\noindent
The resulting code evaluates group-by queries using a Hash Aggregate
strategy, where \verb!tmp! acts as an associative array or hash table.

\subsection{Transforming Multi-block Queries To Single-block Queries}
In traditional query optimizers, subqueries are considered to be
``multi-block queries''. The main query and its subquery are separate
blocks. Transformations exist to turn multi-block queries to single-block
queries. Within the \emph{forelem} framework this is accomplished using the
Inline transformation.

\subsection{Run-Time Optimization of Queries}
An important properly of traditional database systems is that query planning
is done when a query is submitted and statistics about the data that is
currently stored in the database is used in obtaining an efficient query
plan. Because our approach relies heavily on the use of the compiler-based
techniques, one can have the impression that once a query has been optimized
it is static and no changes are made to the optimized query in response to
changes of the stored dataset. For highly dynamic datasets this would put
our approach at a serious disadvantage. However, in the last decades many
techniques have been developed for dynamic code optimization. Examples of
these techniques are Just-In-Time compilation (JIT), multi-version codes and
trace scheduling. By exploiting these techniques, queries optimized with our
approach can still be further optimized at run-time.

As a concrete example, we will discuss how trace scheduling could be used.
In trace scheduling~\cite{fisher-1981,fisher-1984}, all possible paths
through multiple basic blocks (traces) are generated, and code is generated
for each of these paths. In this particular context, this technique is used
to find out what conditions the tuples are tested for and what data is
retrieved from each table in case a condition is satisfied. For conditions
that are tested most frequently or multiple times, according to the traces,
an index set can be generated at run-time before the execution of the
\emph{forelem} loop. Even more powerful is the capability of collecting
traces at run-time and using these in subsequent recompilations of the
query. With this run-time information, better selections can be made for
what index sets to generate at run-time, or to decide to keep persistent
copies of certain index sets updated on disk to avoid recreating the index
set at run-time every time it is needed by a query.

\subsection{Summary}
The correspondences between traditional query optimization and compiler
optimizations are summarized in Table~\ref{tab:query-vs-compiler}.

\begin{table}
\caption{An overview of correspondences between query optimizations and
compiler optimizations.}
\label{tab:query-vs-compiler}
\begin{tabular}{|l|p{5cm}|}
\hline
\textbf{Query Optimization} & \textbf{Compiler Optimizations} \\
\hline
Join Reordering & Loop Interchange \\
\hline
Selection Pushing & Loop Invariant Code Motion \\
\hline
Nested Loops Join & 1. Move all conditions to if-statement in inner loop
body. \\
 & 2. Loop-Independent Materialization of all loops that compose the join.  \\
\hline
%%%
Block Nested Loops Join & 1. Index set partitioning of outer loop. \\
 & 2. Loop Blocking using the partitioned index set. \\
 & 3. Loop Interchange \\
 & 4. Index Set Materialization \\
 & 5. Loop-Independent Materialization of condition-less loop. \\
\hline
Index Nested Loops & 1. Loop Interchange such that available index sets on
disk are used. \\
 & 2. Materialization \\
\hline
Hash Join  &  1. Partition both tables. \\
 & 2. Loop Blocking based on this partitioning. \\
 & 3. Index Set Materialization \\
 & 4. Loop-Independent Materialization of condition-less loop \\
\hline
Sorted Aggregation & 1. Materialize temporary table to sorted array. \\
 & 2. Materialize index set according to sorted array. \\
 & 3. Eliminate redundant iteration. \\
 & 4. Table Propagation \\
 & 5. Dead Code Elimination \\
\hline
Hash Aggregation  & 1. Inline aggregate functions. \\
 & 2. Iteration Space Expansion \\
 & 3. Loop Invariant Code Motion \\
\hline
Multi-block to & 1. Inline subqueries. \\
single-block & 2. Loop Invariant Code Motion. \\
\hline
\end{tabular}
\end{table}

%% Section: Strategy
\section{Optimization and Code Generation Strategies}
\label{sec:strategy}
In order to successfully optimize \emph{forelem} loop nests using the
transformations described in Section~\ref{sec:transformations}, a strategy
is needed that determines in which order to perform the transformations on
the \emph{forelem} loop nests. The \emph{forelem} framework uses the
following algorithm to apply the transformations:

\begin{enumerate}
\item \textbf{Inline}. This inlines subqueries, so that
these can be considered in combination with the calling context.

\item \textbf{Loop Interchange, Loop Invariant Code Motion}. Loops are
reordered such that as many conditions as possible are tested in the
outermost loops. Priority is given to move conditions that test
against a constant value to the outermost loop.

\item \textbf{Iteration Space Expansion}. Opportunities for the application
of Iteration Space Expansion are looked for.  An example of such an
opportunity is a loop iterating an index set with a condition on a
field, of which the body computes an aggregate function. Iteration
Space Expansion is followed by Loop Invariant Code Motion, because
the loop computing the aggregate function is often made loop
invariant by the Iteration Space Expansion transformation. Iteration
Space Expansion is not applied on loops iterating temporary tables.

\item \textbf{Table Propagation}. Through the application of Table
Propagation, preparations are made for the elimination of
unnecessary temporary tables.

\item \textbf{Dead Code Elimination}. At this stage, any loop that computes
unused results is removed.

\item \textbf{Index Set Materialization}. As has been described in Section~
\ref{sec:materialization} for each index set a choice has to be made whether
no index set is generated (implying the table to be fully iterated each time
the index set is used), or to materialize an index set. The subsection on
Materialization Strategy below discusses how this choice is made. For
materialized index sets, opportunities are sought for the application of the
Index Set Pruning and Index Set Combination transformations.
From this step, implementations of the relational join operator
such as Block Nested Loops Join and Hash Join will follow automatically.

\item \textbf{Concretization}. As a final step Concretization takes place.
During this phase it is determined in what format to store the database
tables and what data structures to use to store index sets generated at
run-time.
\end{enumerate}

Experiments have been conducted with the queries from the TPC-H
benchmark~\cite{tpc-h}. The different transformations that have been applied
to each TPC-H query during the \emph{forelem} optimization phase are shown
in Table~\ref{tab:tpc-transformations}.

\begin{table*}
\caption{An overview of the transformations applied to each TPC-H query, in
the order of application, and the used index set concretizations. The
abbreviation LICM stands for Loop Invariant Code Motion.}
\label{tab:tpc-transformations}
\small
\begin{tabular}{|c|l|}
\hline
\textbf{Query \#} & \textbf{Applied Transformations} \\
\hline
1 & Table Propagation, Iteration Space Expansion, LICM,
  Dead Code Elimination. \textbf{Index Set Concretization:} Hash \\
\hline
2 & Inline, Loop Interchange, LICM, Iteration Space Expansion, LICM.
  \textbf{Index Set Concretization:} Hash \\
\hline
3 & Loop Interchange, Iteration Space Expansion, LICM, Index Set Pruning,
  Index Set Combination.\\
  & \textbf{Index Set Concretization:} Hash \\
\hline
4  & Inline, Iteration Space Expansion, Index Set Pruning
   \textbf{Index Set Concretization:} Hash \\
\hline
5  & Loop Interchange, Iteration Space Expansion, LICM, Index Set Pruning,
   Index Set Combination. \\
   & \textbf{Index Set Concretization:} Hash \\
\hline
6  & None\\
\hline
7  & Loop Interchange, LICM, Index Set Pruning, Sorted Aggregation
   \textbf{Index Set Concretization:} Hash \\
\hline
8  & Loop Interchange, LICM, Index Set Pruning, Index Set Combination,
  Sorted Aggregation.\\
   & \textbf{Index Set Concretization:} Hash, array \\
\hline
9  & Loop Interchange, Iteration Space Expansion, LICM, Table Propagation,
   Dead Code Elimination,\\
   & Index Set Pruning, Index Set Combination.
   \textbf{Index Set Concretization:} Hash, array \\
\hline
10 & Loop Interchange, Iteration Space Expansion, Table Propagation,
   Index Set Pruning.\\
   & \textbf{Index Set Concretization:} Hash \\
\hline
11 & Inline, Loop Interchange, Iteration Space Expansion, Loop Fusion, LICM,
   Table Propagation,\\
   & Dead Code Elimination, Index Set Pruning, Index Set Combination. \textbf{Index Set Concretization:} Hash \\
\hline
12 & Loop Interchange, LICM, Index Set Pruning, Index Set Combination
   \textbf{Index Set Concretization:} Hash \\
\hline
13 & Inline, Table Propagation, Dead Code Elimination, Index Set Pruning,
   Sorted Aggregation.\\
   & \textbf{Index Set Concretization:} Balanced tree \\
\hline
14 & Index Set Pruning. \textbf{Index Set Concretization:} Hash \\
\hline
15 & Inline, Loop Interchange, Iteration Space Expansion, LICM,
   Table Propagation, Dead Code Elimination \\
   & \textbf{Index Set Concretization:} Hash \\
\hline
16 & Inline, Loop Interchange, LICM, Table Propagation, Dead Code
Elimination, Index Set Pruning,\\
   & Sorted Aggregation. \textbf{Index Set Concretization:} Hash \\
\hline
17 & Inline, Iteration Space Expansion, LICM, Index Set Pruning \\
   & \textbf{Index Set Concretization:} Array \\
\hline
18 & Inline, Loop Interchange, LICM, Table Propagation, Dead Code Elimination  \\
   & \textbf{Index Set Concretization:} Hash \\
\hline
19 & None. \textbf{Index Set Concretization:} Array \\
\hline
20 & Inline, Iteration Space Expansion, LICM, Index Set Pruning,
   Index Set Combination.\\
   & \textbf{Index Set Concretization:} Hash \\
\hline
21 & Loop Interchange, LICM, Index Set Pruning, Index Set Combination. \\
   & \textbf{Index Set Concretization:} Hash \\
\hline
22 & Inline, LICM, Sorted Aggregation. \textbf{Index Set Concretization:} Hash \\
\hline
\end{tabular}
\normalsize
\end{table*}

Another optimization strategy is to perform a brute-force exploration of the
entire optimization space. This is useful, for example, for queries that are
run many times on changing data so that the costly optimization effort is
worth it. We plan to study brute-force exploration of the optimization
search space in future work.

\subsection*{Materialization Strategy}
Next to the algorithm for the application of transformations on the
\emph{forelem} intermediate representation to optimize queries, a strategy
is defined for the generation of efficient code from the \emph{forelem}
intermediate representation. This strategy is mainly concerned with the
selection of \emph{forelem} loops for which index sets will be materialized
(so that these are generated at
run-time) and the selection of efficient data structures for such index sets.
According to this strategy, index sets that satisfy the following conditions
are \emph{never} generated at run-time:

\begin{enumerate}
\item Index sets without conditions address the full array. In this case,
the index set is not generated, but the loop iterating this index set is
concretized to a simple for loop that iterates the full table of tuples with
subscripts \(i \in [0, \mathit{len})\).
\item Index sets that are iterated by outer loops. These are not generated
because the outer loop is iterated only once.
\item Index sets for very small tables.
\end{enumerate}

\noindent
In all other cases an index set is typically generated. Index sets that are
used in multiple loop nests get priority in being generated. When
materialized index sets are concretized, a suitable data structure has to be
chosen.  For single-dimensional index sets on a field that has a unique
value for each row in the table (one-to-one mapping), often a flat array is
chosen if the key space is known to be small, or otherwise a hash table.
These properties can be known to the code generator because the field was
specified as primary key in the table schema, or the generated code detects
at run-time that the table data satisfies this condition. For index sets
that yield multiple subscripts that are iterated in a loop nest, a balanced
tree is used.

%% Section: Experimental results
\section{Experimental Results}
\label{sec:experiments}
Experiments have been conducted using the queries from the TPC-H
benchmark~\cite{tpc-h}. All queries were parsed into the \emph{forelem}
intermediate representation, optimized using the transformations described
in this paper and C/C++ code has been generated from the optimized AST.
These executables access the database data through
memory-mapped I/O. The execution time of the queries is compared to the
execution time of the same queries as executed by
MonetDB~\cite{monetdb}.

All experiments have been carried out on an Intel Core 2 Quad CPU (Q9450)
clocked at 2.66 GHz with 4 GB of RAM. The software installation consists out
of Ubuntu 10.04.3 LTS (64-bit). The version of MonetDB used is 11.11.11
(Jul2012-SP2), which is the latest version that could be obtained from the
MonetDB website~\cite{monetdb} for use with this operating system.

All queries were run with MonetDB and \emph{forelem}-generated code on a
TPC-H data set of scale factor 1.0. The speedups achieved by achieved by the
\emph{forelem}-optimized queries over MonetDB are shown in
Table~\ref{tab:tpch}. These speedups range from a factor of 1.21
(Q3) to 4.09 (Q15).

MonetDB and the \emph{forelem}-generated code, have also been tested on a
data set with scale factor 10.0. The speedups that are achieved by the
\emph{forelem}-optimized queries over MonetDB are shown in
Table~\ref{tab:tpch10}. Also in this case, the
\emph{forelem}-optimized queries perform significantly better than MonetDB
in all cases. The speedups range from a factor of 1.03 (Q8) to 15.6 (Q18).

\begin{table}
\centering
\caption{Speedup of the execution time of TPC-H queries optimized with the
\emph{forelem} framework compared to MonetDB on a data set of scale factor
1.0.}
\label{tab:tpch}
\begin{tabular}{l|c||l|c|}
\textbf{Query} & \textbf{Speedup} & \textbf{Query} & \textbf{Speedup} \\
\hline
Q1  & 2.84 & Q12 & 2.67  \\
Q2  & 1.65 & Q13 & 1.94  \\
Q3  & 1.21 & Q14 & 2.08  \\
Q4  & 1.39 & Q15 & 4.09  \\
Q5  & 1.33 & Q16 & 1.37  \\
Q6  & 3.64 & Q17 & 3.22  \\
Q7  & 2.79 & Q18 & 2.68  \\
Q8  & 1.63 & Q19 & 1.71  \\
Q9  & 1.28 & Q20 & 1.38  \\
Q10 & 1.62 & Q21 & 1.55  \\
Q11 & 1.98 & Q22 & 1.39  \\
\hline
\end{tabular}
\end{table}
\begin{table}
\centering
\caption{Speedup of the execution time of TPC-H queries optimized with the
\emph{forelem} framework compared to MonetDB on a data set of scale
factor 10.0.}
\label{tab:tpch10}
\begin{tabular}{l|c||l|c|}
\textbf{Query} & \textbf{Speedup} & \textbf{Query} & \textbf{Speedup} \\
\hline
Q1  & 8.81 & Q12 & 6.33  \\
Q2  & 3.06 & Q13 & 1.34  \\
Q3  & 5.00 & Q14 & 1.74  \\
Q4  & 5.19 & Q15 & 1.38  \\
Q5  & 1.41 & Q16 & 1.16  \\
Q6  & 4.00 & Q17 & 1.47  \\
Q7  & 3.00 & Q18 & 15.61  \\
Q8  & 1.03 & Q19 & 3.56  \\
Q9  & 2.81 & Q20 & 1.11  \\
Q10 & 4.54 & Q21 & 3.40  \\
Q11 & 2.27 & Q22 & 3.21  \\
\hline
\end{tabular}
\end{table}

%% Section: Implementation
\section{Implementation of the Forelem Framework}
\label{sec:forelem-implementation}
The \emph{forelem} framework has been developed as a generic library,
\emph{libforelem}, to be able to support different programming languages and
data access frameworks. This library is capable of creating and
manipulating \emph{forelem} loop nests, by representing these using an
internal Abstract Syntax Tree (AST).

Different applications can make use of \emph{libforelem} to create and
manipulate \emph{forelem} ASTs. For example, to support the vertical
integration of database applications, the \emph{libforelem} library is
capable of parsing a given SQL statement into a \emph{forelem} AST.  On the
AST, various analyses and transformations can be applied, many of which are
implementations of traditional compiler (loop) transformations that function
on the \emph{forelem} AST. An abstract code generation interface is present
in the library to generate code from any \emph{forelem} AST. Currently, the
output of C/C++ code and algebraic \emph{forelem} is supported. However, the
use of \emph{forelem} loops is not restricted to C/C++ and other languages
can be supported by implementing the abstract code generation interface.

Typically in the optimization process, the code generator is called when
optimization on the \emph{forelem}-loop level has completed. In the case
C/C++ code is to be generated from the optimized \emph{forelem} AST, the
\emph{forelem} loops are translated to C \emph{for} loops that iterate index
sets and access subscripts of plain C arrays. In the C code, an index set is
a generic interface and the exact data structure of the index set is opaque.
As has been described in Section~\ref{sec:strategy}, the
optimization process will generate a data storage format for each index set
as part of the optimization process. Since materialization of index sets is
not obligatory, the optimization process might also decide to not generate
an index set, but rather to test the conditions during iteration of the
table.

Note that \emph{forelem} loops are only used by the compiler tooling and
are never visible to the end user. The general nature of the \emph{forelem}
frameworks allows for its usage with other problems. Other parsers that take
a certain language as input and produce \emph{forelem} loops can be
developed next to the SQL parser, so that other problem domains can be
supported.
\section{Related Work}
\label{sec:related-work}
Several methodologies to increase the performance of database query
evaluation with the use of optimizing compiler technology have been
described in the past. A strategy to transform entire queries to executable
code is described in~\cite{krikellas-2010}. The technology, called
``holistic query evaluation'', works by transforming a query evaluation plan
into source code, based on code templates, and compiling this into a shared
library using an aggressively optimizing compiler. The shared library is
then linked into the database server for processing. Although significant
speed-ups over traditional and currently-emerging database systems are
achieved, this approach remains fixated to traditional query planning
techniques to optimize the query. Optimizing compiler technology is used to
compile a translation of the query plan into C/C++ code through the use of code
templates into efficient executable code, contrary to our approach which
uses compiler technology to replace the traditional query planning.

The UltraLite system, described in~\cite{ultralite-2002}, follows a similar
approach and compiles queries used in an embedded SQL code to C code. This
is achieved by sending the query to a host database server, which parses and
optimizes the query using traditional techniques. An execution plan is
returned which is used to generate the C code.

In~\cite{neumann-2011} a data-centric approach to query compilation is
described. SQL queries are translated to an algebraic query execution plan
from which LLVM bitcode is generated. The query in LLVM bitcode is then
executed using the optimizing JIT compiler included with LLVM. The
translation of the query plan to LLVM bitcode is performed with a
data-centric approach, in the sense that care is taken to keep data in CPU
registers as long as possible for optimal performance, instead of
maintaining clear operator boundaries. This technique results in very
efficient query codes.

Compiler optimizations have also been used to take on the problem of
multi-query optimization. In~\cite{andrade-2003} an approach is demonstrated
where queries are written as imperative loops, on which compiler
optimization strategies are applied. The use of loop fusion, common
subexpression elimination and dead code elimination is described. This work
is tailored towards a certain class of analysis queries and not to
generic queries as is the case with \emph{forelem}. Furthermore,
the Loop Fusion transformation described in the paper works by detecting
multidimensional overlap, so, the strategy of Loop Fusion is used, but not
an exact mapping of the traditional Loop Fusion optimization which is the
aim for the \emph{forelem} framework.

A different approach to multi-query optimization is described
in~\cite{kang-1994}, where optimization techniques are applied to the
``algo\-rithm-level'' of a database program. In the algorithm-level, a query
is represented as a sequence of algorithms, e.g. selection, join, that
should be performed to compute the query results. The exact implementation
of the algorithms is not made explicit at this level. As a
consequence, knowledge is required about the implementation of algorithms
that can appear in the representation by the optimizer in order to be able
to carry out optimizations. In \emph{forelem}, all operations are
transformed
to iterations of arrays and the focus is on performing optimizations on a
series of loop nests. A notable exception are the sorting and duplicate
elimination operations however, as these are either specified as a modifying
operation of a result set or as a condition to an index set. Array
processing is sped up by selecting effective index sets. Index sets are
either generated statically or specific algorithms are selected for run-time
index set generation at the time executable code is generated from
\emph{forelem} loop nests.

A specific class of programming languages exists that include the ability to
iterate through sets. These languages are known as Database Programming
Languages (DBPLs).  For these languages, compile-time optimizations similar
to relational transformations like join reordering have been
described~\cite{lieuwen-1992}. These transformations make standard
trans\-formation-based compilers capable of optimizing iterations over sets
that correspond to joins. This was later extended to include transformations
that enable the parallelization of loops in DBPLs~\cite{lieuwen-1998}.

There are a number of important differences with the \emph{forelem}
framework. DBPLs were meant as programming languages to be used by an end
user, whereas \emph{forelem} loops are only an intermediate representation
used to be used by code optimization backends. This way, the \emph{forelem}
framework is capable of handling different combinations of application
programming languages and database statement expressions. Additionally,
DBPLs, such as for example O++ which is discussed in the cited papers, use a
run-time system for the iteration and manipulation of sets.  \emph{forelem}
loops can be immediately lowered to low-level C codes that iterate over
arrays that enable further low-level compiler optimizations.  Despite of
that, several of the techniques presented in these papers could be
implemented in the \emph{forelem} framework in the future.
\section{Conclusions}
\label{sec:conclusions}
In this paper, the optimization of database queries using compiler
transformations has been described. This optimization process is carried out
in the \emph{forelem} framework. The \emph{forelem} framework provides an
intermediate representation to which queries can be naturally transformed
and on which compiler transformations can be applied to optimize the loop
nest, contrary to the traditional optimization of queries by the generation
of an efficient query execution plan. Compiler transformations that are
currently implemented within the \emph{forelem} framework were illustrated.
Using these transformations, combined with Materialization techniques,
established implementations of relational join operators such as Hash Join
can be derived automatically. Finally, strategies for the application of
these transformations were discussed to be able to optimize query codes to
achieve higher performance than contemporary database systems.

Experiments using the queries from the TPC-H benchmark show that
using compiler transformations implemented within the \emph{forelem}
framework the queries could be optimized to perform significantly better
than contemporary database systems. An average improvement was shown of a
factor of 3 and in certain cases a speedup of up to a factor of 15.
\bibliographystyle{IEEEtran}

\begin{thebibliography}{10}
\providecommand{\url}[1]{#1}
\csname url@samestyle\endcsname
\providecommand{\newblock}{\relax}
\providecommand{\bibinfo}[2]{#2}
\providecommand{\BIBentrySTDinterwordspacing}{\spaceskip=0pt\relax}
\providecommand{\BIBentryALTinterwordstretchfactor}{4}
\providecommand{\BIBentryALTinterwordspacing}{\spaceskip=\fontdimen2\font plus
\BIBentryALTinterwordstretchfactor\fontdimen3\font minus
  \fontdimen4\font\relax}
\providecommand{\BIBforeignlanguage}[2]{{%
\expandafter\ifx\csname l@#1\endcsname\relax
\typeout{** WARNING: IEEEtran.bst: No hyphenation pattern has been}%
\typeout{** loaded for the language `#1'. Using the pattern for}%
\typeout{** the default language instead.}%
\else
\language=\csname l@#1\endcsname
\fi
#2}}
\providecommand{\BIBdecl}{\relax}
\BIBdecl

\bibitem{rietveld-2015}
K.~F.~D. Rietveld and H.~A.~G. Wijshoff, ``Reducing layered database
  applications to their essence through vertical integration,'' \emph{ACM
  Trans. Database Syst.}, vol.~40, no.~3, pp. 18:1--18:39, 2015.

\bibitem{tpc-h}
{Transaction Processing Performance Council}, ``{TPC-H},''
  \newline\url{http://tpc.org/tpch/default.asp}, May 2009.

\bibitem{krikellas-2010}
K.~Krikellas, S.~Viglas, and M.~Cintra, ``{Generating code for holistic query
  evaluation},'' in \emph{{ICDE}}, 2010, pp. 613--624.

\bibitem{neumann-2011}
T.~Neumann, ``{Efficiently compiling efficient query plans for modern
  hardware},'' \emph{Proc. VLDB Endow.}, vol.~4, pp. 539--550, Jun. 2011.

\bibitem{maier-1990}
D.~Maier, \emph{{Representing database programs as objects}}.\hskip 1em plus
  0.5em minus 0.4em\relax New York, NY, USA: ACM, 1990, pp. 377--386.

\bibitem{kennedy-1994}
K.~Kennedy and K.~McKinley, ``Maximizing loop parallelism and improving data
  locality via loop fusion and distribution,'' in \emph{Languages and Compilers
  for Parallel Computing}, ser. Lecture Notes in Computer Science, U.~Banerjee,
  D.~Gelernter, A.~Nicolau, and D.~Padua, Eds.\hskip 1em plus 0.5em minus
  0.4em\relax Springer Berlin / Heidelberg, 1994, vol. 768, pp. 301--320.

\bibitem{kuck-1981}
D.~J. Kuck, R.~H. Kuhn, D.~A. Padua, B.~Leasure, and M.~Wolfe, ``Dependence
  graphs and compiler optimizations,'' in \emph{Proceedings of the 8th ACM
  SIGPLAN-SIGACT symposium on Principles of programming languages}, ser. POPL
  '81.\hskip 1em plus 0.5em minus 0.4em\relax New York, NY, USA: ACM, 1981, pp.
  207--218.

\bibitem{allen-phd}
A.~J. R., \emph{Dependence Analysis for Subscripted Variables and its
  Applications to Program Transformations}.\hskip 1em plus 0.5em minus
  0.4em\relax PhD Dissertation, Rice University, 1983.

\bibitem{allen-1987}
R.~Allen and K.~Kennedy, ``Automatic translation of fortran programs to vector
  form,'' \emph{ACM Trans. Program. Lang. Syst.}, vol.~9, pp. 491--542, October
  1987.

\bibitem{zima-1991}
H.~Zima and B.~Chapman, \emph{{Supercompilers for parallel and vector
  computers}}.\hskip 1em plus 0.5em minus 0.4em\relax New York, NY, USA: ACM,
  1991.

\bibitem{allen-1976}
F.~E. Allen and J.~Cocke, ``A program data flow analysis procedure,''
  \emph{Commun. ACM}, vol.~19, no.~3, pp. 137--, Mar. 1976.

\bibitem{kennedy-1981}
K.~Kennedy, \emph{{A survey of data flow analysis techniques}}.\hskip 1em plus
  0.5em minus 0.4em\relax Englewood Cliffs NJ: Prentice-Hall, 1981, pp. 5--54.

\bibitem{rietveld-2013-cpc}
K.~F.~D. Rietveld and H.~A.~G. Wijshoff, ``{Forelem: A Versatile Optimization
  Framework For Tuple-Based Computations},'' in \emph{{CPC 2013: 17th Workshop
  on Compilers for Parallel Computing}}, July 2013.

\bibitem{lam-1991}
M.~D. Lam, E.~E. Rothberg, and M.~E. Wolf, ``The cache performance and
  optimizations of blocked algorithms,'' \emph{SIGARCH Comput. Archit. News},
  vol.~19, no.~2, pp. 63--74, Apr. 1991.

\bibitem{temam-1993}
O.~Temam, E.~Granston, and W.~Jalby, ``To copy or not to copy: A compile-time
  technique for assessing when data copying should be used to eliminate cache
  conflicts,'' in \emph{Supercomputing '93. Proceedings}, 1993, pp. 410--419.

\bibitem{ramakrishnan-2003}
R.~Ramakrishnan and J.~Gehrke, \emph{Database management systems (3.
  ed.)}.\hskip 1em plus 0.5em minus 0.4em\relax McGraw-Hill, 2003.

\bibitem{chaudhuri-1998}
S.~Chaudhuri, ``An overview of query optimization in relational systems,'' in
  \emph{Proceedings of the seventeenth ACM SIGACT-SIGMOD-SIGART symposium on
  Principles of database systems}.\hskip 1em plus 0.5em minus 0.4em\relax ACM,
  1998, pp. 34--43.

\bibitem{knijnenburg-2003}
P.~Knijnenburg, T.~Kisuki, and M.~O'Boyle,
  ``\BIBforeignlanguage{English}{Combined selection of tile sizes and unroll
  factors using iterative compilation},''
  \emph{\BIBforeignlanguage{English}{The Journal of Supercomputing}}, vol.~24,
  no.~1, pp. 43--67, 2003.

\bibitem{fursin-2005}
G.~Fursin, M.~O'Boyle, and P.~Knijnenburg, ``Evaluating iterative
  compilation,'' in \emph{Languages and Compilers for Parallel Computing}, ser.
  Lecture Notes in Computer Science, B.~Pugh and C.-W. Tseng, Eds.\hskip 1em
  plus 0.5em minus 0.4em\relax Springer Berlin Heidelberg, 2005, vol. 2481, pp.
  362--376.

\bibitem{fisher-1981}
J.~Fisher, ``Trace scheduling: A technique for global microcode compaction,''
  \emph{IEEE Transactions on Computers}, vol.~30, no.~7, pp. 478--490, 1981.

\bibitem{fisher-1984}
J.~A. Fisher, J.~R. Ellis, J.~C. Ruttenberg, and A.~Nicolau, ``Parallel
  processing: a smart compiler and a dumb machine,'' \emph{SIGPLAN Not.},
  vol.~19, no.~6, pp. 37--47, Jun. 1984.

\bibitem{monetdb}
\BIBentryALTinterwordspacing
{MonetDB Project}, ``{MonetDB},'' February 2013. [Online]. Available:
  \url{http://www.monetdb.org/}
\BIBentrySTDinterwordspacing

\bibitem{ultralite-2002}
D.~P. Yach, J.~D. Graham, and A.~F. Scian, ``Database system with methodology
  for accessing a database from portabl e devices,'' US Patent \#6341288, Jan
  2002.

\bibitem{andrade-2003}
H.~Andrade, S.~Aryangat, T.~M. Kur\c{c}, J.~H. Saltz, and A.~Sussman,
  ``Efficient execution of multi-query data analysis batches using compiler
  optimization strategies,'' in \emph{LCPC}, 2003, pp. 509--524.

\bibitem{kang-1994}
M.~H. Kang, H.~G. Dietz, and B.~K. Bhargava, ``Multiple-query optimization at
  algorithm-level,'' \emph{Data Knowl. Eng.}, vol.~14, no.~1, pp. 57--75, 1994.

\bibitem{lieuwen-1992}
D.~F. Lieuwen and D.~J. DeWitt, ``{A Transformation-Based Approach to
  Optimizing Loops in Database Programming Languages},'' in \emph{{SIGMOD
  Conference}}, 1992, pp. 91--100.

\bibitem{lieuwen-1998}
D.~F. Lieuwen, ``{Parallelizing Loops in Database Programming Languages},'' in
  \emph{{ICDE}}, 1998, pp. 86--93.

\end{thebibliography}


%
\end{document}